\documentclass[aps,prd,showpacs,twocolumn,superscriptaddress]{revtex4-1}
\usepackage{amsmath}
\usepackage{amsfonts}
\usepackage{amssymb}
\usepackage{graphicx}
\usepackage{dcolumn}
\usepackage{bm}
\usepackage{amsthm}
\usepackage[caption=false]{subfig}
\usepackage{natbib}

\begin{document}
\title{Accelerated Expansion of the Universe based on Emergence of Space
and Thermodynamics of the Horizon}

\author{Fei-Quan Tu}
 \email{fqtuzju@foxmail.com}
 \affiliation{School of Physics and Electronic Science, Zunyi Normal University, Zunyi 563006, China}
\author{Yi-Xin Chen}
 \email{yxchen@zimp.zju.edu.cn}
 \affiliation{Zhejiang Institute of Modern Physics, Zhejiang University, Hangzhou 310027, China}
\author{Bin Sun}
 \email{sunbin@hnu.edu.cn}
 \affiliation{School of Physics and Electronic Science, Zunyi Normal University, Zunyi 563006, China}
\author{You-Chang Yang}
 \email{youcyang@163.com}
 \affiliation{School of Physics and Electronic Science, Zunyi Normal University, Zunyi 563006, China}

\begin{abstract}
Researches in the several decades have shown that the dynamics of
gravity is closely related to thermodynamics of the horizon. In this
paper, we derive the Friedmann acceleration equation based on the
idea of ``emergence of space'' and thermodynamics of the Hubble
horizon whose temperature is obtained from the unified first law of
thermodynamics. Then we derive another evolution equation of the universe
based on the energy balance relation $\rho V_{H}=TS$.
Combining the two evolution equations and the equation
of state of the cosmic matter, we obtain the evolution solutions of
the FRW universe. We find that the solutions obtained by us include
the solutions obtained in the standard general relativity(GR) theory.
Therefore, it is more general to describe the evolution of the
universe in the thermodynamic way.
\end{abstract}

\maketitle

\section*{1. Introduction}
Numerous astronomical observations tell us that the universe is in
accelerated expansion\cite{key-1,key-2,key-3,key-4,key-5,key-6}.
In order to explain the accelerated expansion, a cosmological constant
is usually added to the Einstein field equation. In particular, $\Lambda$
cold dark matter ($\Lambda CDM$) model with 4\% usual matter, 23\%
cold dark matter and 73\% dark energy describes the accelerated expansion
of the universe well. In addition, the accelerated expansion can also
been described by modifying the geometrical part of the field equation
(For example, $f(R)$ gravity\cite{key-7,key-8} and Lanczos-Lovelock
gravity).

Recently, Padmanabhan\cite{key-9,key-10} have suggested that the
difference between the number of the surface degrees of freedom $N_{sur}$
and the number of the bulk degrees of freedom $N_{bulk}$ in a region
of space drives the accelerated expansion of the universe through
a simple equation $\frac{dV}{dt}=L_{p}^{2}(N_{sur}-N_{bulk})$,
where $V$ is the Hubble volume and $L_{p}$ is the Planck length.
The standard Friedmann equation of the FRW universe can be derived
in this emergence of space scenario. Subsequently, emergent perspective
of gravity was further investigated by many researchers\cite{key-11,key-12,key-13}.

On the other hand, it is interesting and meaningful to study cosmology
from the point of view of thermodynamics. In 1995, Jacobson\cite{key-14}
argued that the Einstein equation can be derived from the proportionality
of entropy and horizon area together with the Clausius relation $\delta Q=TdS$
with $\delta Q$ and $T$ interpreted as the energy flux and Unruh
temperature seen by an accelerated observer just inside the horizon.
He also pointed out that the Einstein equation is an equation of state.
His paper revealed that thermodynamics of spacetime is closely related
to dynamics of gravity. Since then, the relationship between thermodynamics
of spacetime and dynamics of gravity have been investigated widely.
It has been shown that the field equations are equivalent to the thermodynamic
identity $TdS=dE+PdV$ on the horizons in a very wide class of gravitational
theories, such as on the static spherically symmetric horizons\cite{key-15},
the stationary axisymmetric horizons and evolving spherically symmetric
horizons\cite{key-16} in Einstein gravity, static spherically symmetric
horizons\cite{key-17} and dynamical apparent horizons\cite{key-18}
in Lanczos-Lovelock gravity, three dimensional BTZ black hole horizons\cite{key-19,key-20}
etc.

Einstein's equations are equivalent to the unified first law of thermodynamics
for the dynamical black hole when the notion of trapping horizon is
introduced in the GR theory\cite{key-21,key-22,key-23,key-24}. Inspired
by this conclusion, our universe may be considered as a non-stationary
gravitational system\cite{key-25}. Thus we can obtain the temperature
of Hubble horizon of the universe based on the unified first law.
The advantage of obtaining the horizon temperature in this way is
that the temperature has a definite physical origin and a clear mathematical
expression.

In many references (e.g. Refs.\cite{key-11,key-12,key-13,key-26})
which describe the evolution of the FRW universe, the temperature
of Hubble horizon is usually assumed to be $T=H/2\pi$. However, the
form of temperature cannot be obtained from an elegant physical principle.
In this paper, we consider the accelerated expansion of the universe
in the late time based on emergence of space and thermodynamics of
the Hubble horizon. We employ the temperature obtained from the unified
first law of thermodynamics as the temperature of the horizon. Furthermore,
we obtain the number of modified bulk degrees of freedom and get the
corresponding dynamical equations in the FRW universe based on emergence
of space. Then we obtain another evolution equation of the universe
based on the energy balance relation $\rho V_{H}=TS$, where $S$
is the entropy associated with the area of the Hubble sphere and $TS$
is the heat energy of the boundary surface. Combining the two evolution
equations and the equation of state of the cosmic matter, we determine
the evolution laws of the universe. By analyzing the solutions of
the evolution laws, we find the solutions obtained by us include the
solutions obtained in the standard general relativity(GR) theory.
Therefore, it is more general to describe the evolution of the
universe in the thermodynamic way. Next we analyze the solutions in
order for the completeness of the discussion. Finally, we make some
comments on the alternative perspective for the evolution of the universe.

The present paper is organized as follows. In section 2, we give a
simple review about Padmanabhan's work and obtain the temperature
of Hubble horizon based on a good motivation. Furthermore, we derive
the modified Friedmann equations of the FRW universe based on emergence
of space and thermodynamics of the Hubble horizon. In section 3, we
obtain the solutions of the Friedmann equations of the FRW universe
and analyze their physical meanings. In section 4, some comments on
the alternative perspective for the evolution of the universe are
made. Finally, we present our conclusions. In this paper, we choose
the natural units $c=\hbar=1$ and the current Hubble constant $H_{0}=70km\cdot s^{-1}\cdot Mpc^{-1}$.

\section*{2. Modified Friedmann equations }

Our universe is homogeneous and isotropic according to astronomical
observations and can be described by the FRW metric
\begin{equation}
ds^{2}=-dt^{2}+a^{2}(t)\left(\frac{dr^{2}}{1-kr^{2}}+r^{2}d\Omega^{2}\right)=h_{ab}dx^{a}dx^{b}+R^{2}d\Omega^{2},
\end{equation}
where $R=a(t)r$ is the comoving radius, $h_{ab}=diag\left(-1,\;\frac{a^{2}}{1-kr^{2}}\right)$is
the metric of 2-spacetime ($x^{0}=t,\; x^{1}=r$) and $k=0,\;\pm1$
denotes the curvature scalar. Solving the Einstein field equation
under this metric, we can obtain the properties of the evolution of
the FRW universe.

However, Padmanabhan\cite{key-9,key-10} came up with a very interesting
idea that cosmic space is emergent as cosmic time progresses. First
of all, a de Sitter space with the Hubble radius $H^{-1}$ is considered.
The temperature of the horizon of such a space is $T=H/2\pi$. He
defined the notion of surface and bulk degrees of freedom in a spherical
space of radius $H^{-1}$. The number of the surface degrees of freedom
is defined by
\begin{equation}
N_{sur}\equiv\frac{4\pi}{H^{2}L_{p}^{2}},
\end{equation}
where $L_{p}$ is the Planck constant. The number of the bulk degrees
of freedom $N_{bulk}$ is given by the equipartition theorem
\begin{equation}
N_{bulk}=\frac{|E|}{(1/2)k_{B}T}=-\frac{2(\rho+3p)V}{k_{B}T},
\end{equation}
where $|E|$ is the magnitude of Komar energy $|(\rho+3p)|V$, $V$
is Hubble volume $V=4\pi/3H^{3}$. The principle of holographic equipartition
requires
\begin{equation}
N_{sur}=N_{bulk}.
\end{equation}
If the equation of state in the de Sitter space is $p=-\rho$, then
Eq.(4) can be reduced to $H^{2}=8\pi L_{p}^{2}\rho/3$ which is the
standard result of the evolution of the de Sitter universe in the
GR theory. This result also shows that the evolution of the de Sitter
space obeys the principle of holographic equipartition.

Furthermore, our universe is asymptotically de Sitter, so he thought
that the difference between $N_{sur}$ and $N_{bulk}$ drives the
universe towards holographic equipartition and the evolution of the
universe is dominated by
\begin{equation}
\frac{dV}{dt}=L_{p}^{2}\left(N_{sur}-N_{bulk}\right).
\end{equation}
Thus substituting Eq.(2) and Eq.(3) into Eq.(5) and using the relations
$V=4\pi/3H^{3}$, $T=H/2\pi$, we obtain the following relation
\begin{equation}
\frac{\ddot{a}}{a}=H^{2}+\dot{H}=-\frac{4\pi L_{p}^{2}}{3}(\rho+3p).
\end{equation}
This result is the same as that obtained by solving the Einstein field
equation.

In Refs.\cite{key-27,key-28}, the authors derived the temperature
of the horizon of the de Sitter space $T=H/2\pi$ by using the field
theory where $H^{-1}$ is the radius of the de Sitter space. However,
our universe is asymptotically de Sitter rather than purely de Sitter,
so the temperature of the horizon of the universe may not necessarily
be expressed as $H/2\pi$. Thus the following two questions arise:
What is the expression of the temperature on the horizon of the FRW
universe? How can we obtain this expression? Fortunately, we can obtain
the expression of the temperature on the horizon of the FRW universe
from an elegant physical principle.

The above FRW metric(Eq.(1)) can be written in the double-null form\cite{key-29,key-30}
as
\begin{equation}
ds^{2}=-2d\xi^{+}d\xi^{-}+R^{2}d\Omega_{2}^{2},
\end{equation}
where $\partial_{\pm}=\frac{\partial}{\partial\xi^{\pm}}=-\sqrt{2}\left(\frac{\partial}{\partial t}\pm\frac{\sqrt{1-ar^{2}}}{a}\frac{\partial}{\partial r}\right)$
are future pointing null vectors.

The trapping horizon is defined as $\partial_{+}R|_{R=R_{T}}=0$,
which gives

\begin{equation}
R_{T}=\frac{1}{\sqrt{H^{2}+\frac{k}{a^{2}}}}=R_{A},
\end{equation}
where $R_{A}$ is the apparent horizon. The surface gravity is defined
as
\begin{equation}
\kappa=\frac{1}{2\sqrt{-h}}\partial_{a}\left(\sqrt{-h}h^{ab}\partial_{b}R\right),
\end{equation}
so we can get
\begin{equation}
\kappa=-\left(1+\frac{\dot{H}}{2H^{2}}\right)H
\end{equation}
for the apparent horizon of the flat universe. Here we would like
to make some explanations on why we choose the flat universe ($k=0$):
(1). Our universe is flat according to astronomical observations;
(2). We have chosen $1/H$ as the radius of the horizon when we obtain
the dynamical equation of the cosmic evolution, so we also choose
Hubble horizon here in order for the consistency.

According to the relation between the surface gravity and the temperature,
we obtain the temperature of the horizon\cite{key-29,key-30,key-31}
\begin{equation}
T=\frac{|\kappa|}{2\pi}=\frac{H}{2\pi}\left(1+\frac{\dot{H}}{2H^{2}}\right).
\end{equation}

It is emphasized that the definition of the surface
gravity Eq.(9) and the temperature on the horizon Eq.(11) are from
the unified first law. The unified first law is $dE=A\Psi+WdV$ where
$E$ is Misner-Sharp energy, $\Psi_{a}$ is energy-supply, $W$ is
work, $A$ and $V$ are the area and volume of a sphere. One can obtain
the projection relation $<dE,\;\xi>=\frac{\kappa}{8\pi G}<dA,\;\xi>+<WdV,\;\xi>$
where $\xi$ is a projection vector and $\kappa$ is defined as Eq.(9)
when projecting the unified first law along the inner trapping horizon
of the FRW universe in the GR theory\cite{key-29}. On the inner trapping
horizon, the projection relation can be expressed as $<A\Psi,\;\xi>=\frac{\kappa}{8\pi G}<dA,\;\xi>$.
Note that $A\Psi$ when projected on the trapping horizon gives the
heat flow $\delta Q$ , so the projection relation implies the Clausius
relation $\delta Q=TdS$ if we employ that $T=|\kappa|/2\pi$ and
$S=A/4G$. Only when the temperature (or the surface gravity) is defined
in this way can the unified first law imply the Clausius relation.
Therefore, we can say that the expression (11) of the temperature
on the horizon of the FRW universe obtained by us originates from
the fundamental physical relation, i.e. the unified first law.

Substituting Eq.(2), Eq.(3) and Eq.(11) into Eq.(5), we get
\begin{equation}
\frac{\ddot{a}}{a}=H^{2}+\dot{H}=-\frac{4\pi L_{p}^{2}}{3}(\rho+3p)\left(1+\frac{\dot{H}}{2H^{2}}\right)^{-1}.
\end{equation}
This is the evolution equation of the FRW universe when we employ
the idea of emergence of space and reconsider the temperature of the
Hubble horizon. Comparing Eq.(12) with the standard Friedmann acceleration
equation $\frac{\ddot{a}}{a}=H^{2}+\dot{H}=-\frac{4\pi L_{p}^{2}}{3}(\rho+3p)$
in the GR theory, we find the term about the density of the energy
and pressure has a correction in our model. Furthermore, Eq (12) can
be rewritten as
\begin{equation}
\frac{\ddot{a}}{a}=H^{2}+\dot{H}=-\frac{4\pi L_{p}^{2}}{3}(\rho+3p)-\frac{\dot{H}}{2}-\frac{\dot{H^{2}}}{2H^{2}}.
\end{equation}

In the GR theory, the Friedmann acceleration equation with cosmological
constant $\Lambda$ is $\frac{\ddot{a}}{a}=H^{2}+\dot{H}=-\frac{4\pi L_{p}^{2}}{3}(\rho_{m}+3p_{m})+\frac{\Lambda}{3}$
where the subscript $m$ represents matter and the positive constant
$\Lambda$ is explained as the vacuum energy. Introducing cold dark
matter again, this equation can describe the accelerated expansion
of the universe well. In the model of entropic cosmology\cite{key-32,key-33},
the Friedmann acceleration equation is given as $\frac{\ddot{a}}{a}=H^{2}+\dot{H}=-\frac{4\pi L_{p}^{2}}{3}(\rho_{m}+3p_{m})+C_{H}H^{2}+C_{\dot{H}}\dot{H}$
where the coefficients are bounded by $\frac{3}{2\pi}\lesssim C_{H}\leq1$
and $0\leq C_{\dot{H}}\lesssim\frac{3}{4\pi}$. The reason that the
term $C_{H}H^{2}+C_{\dot{H}}\dot{H}$ is added to the Friedmann acceleration
equation is that the surface term in the gravitational action can
not be ignored. The accelerated expansion of the late universe\cite{key-32}
and the inflation of the early universe\cite{key-33} can be explained
by adding this term. So the general formula of the Friedmann acceleration
equation can be summarized as (See, for example, in Refs.\cite{key-32,key-33,key-34,key-35})
\begin{equation}
\frac{\ddot{a}}{a}=H^{2}+\dot{H}=-\frac{4\pi L_{p}^{2}}{3}(\rho_{m}+3p_{m})+f(H,\;\dot{H}),
\end{equation}
where $f(H,\;\dot{H})$ is a function of $H$ and $\dot{H}$. The
specific forms of Eq.(14) have been used to discuss cosmology widely,
which result in some good results (For example in Refs.\cite{key-34,key-35,key-36}).
However, $\rho$ and $p$ are the effective energy density and pressure
which include the equivalent energy and pressure of the vacuum energy
in our model.

If we compare Eq.(13) with the Friedmann acceleration equation $\frac{\ddot{a}}{a}=H^{2}+\dot{H}=-\frac{4\pi L_{p}^{2}}{3}(\rho_{m}+3p_{m})+\frac{\Lambda}{3}$
in the GR theory, we find that the vacuum energy $\Lambda$ has a
shift $-\frac{3\dot{H}}{2H^{2}}(H^{2}+\dot{H})=-\frac{3\dot{H}}{2H^{2}}\frac{\ddot{a}}{a}$
which changes with time therefore it is not a constant in our model.
Analyzing the shift $-\frac{3\dot{H}}{2H^{2}}\frac{\ddot{a}}{a}$
, we find that the total vacuum energy is always positive as long
as the universe is in accelerated expansion and $\dot{H}<0$ . On
the other hand, the results that the vacuum energy is positive and
the universe is in accelerated expansion are consistent with the
present astronomical observations. Thus the acceleration equation
(13) has a good physical explanation.

Now we would like to make some comments about the Friedmann acceleration
equation Eq.(13). Firstly, the unambiguous derivation of Eq.(13) is
just based on emergence of space and the reasonable temperature of
the Hubble horizon of the FRW universe. In contrast, the model in
Ref.\cite{key-32} is a phenomenological model in which the surface
term is introduced without rigorous derivation. Secondly, the temperature
of the horizon of the de Sitter space have been derived\cite{key-27,key-28},
but the temperature of the Hubble horizon of the FRW universe, which
is not purely de Sitter space, is just assumed to be $T=H/2\pi$ in
many references as we have pointed out. However, we employ the temperature
of the Hubble horizon which can be obtained from the unified
first law\cite{key-29,key-30} in this paper. Thirdly, the specific
forms of $f(H,\;\dot{H})$ in Eq.(14) have been studied widely and
some good results have been obtained. Moreover, our equation is consistent
with astronomical observations. Finally, we would like to stress that
the derivation of the acceleration equation (13) is based on an elegant
principle (the unified first law) rather than an assumption and this
is an advantage of our model.

On the other hand, the Friedmann equation $\rho=3H^{2}/(8\pi G)$
can be rewritten as an energy balance relation\cite{key-37}
\begin{equation}
\rho V_{H}=TS
\end{equation}
in thermodynamic language, where $S=A_{H}/(4L_{p}^{2})=\pi H^{-2}/L_{p}^{2}$
is the entropy associated with the area of the Hubble sphere $V_{H}=\frac{4\pi}{3H^{3}}$
and $TS$ is the heat energy of the boundary surface.

From now on, we take the energy balance relation (15) as the fundamental
equation describing the change of the energy density of cosmic matter.
Substituting Eq.(11) and $S=\pi H^{-2}/L_{p}^{2}$ into the energy
balance relation (15), we obtain
\begin{equation}
H^{2}=\frac{8\pi L_{p}^{2}}{3}\left(1+\frac{\dot{H}}{2H^{2}}\right)^{-1}\rho.
\end{equation}
This equation can be rewritten as
\begin{equation}
H^{2}=\frac{8\pi L_{p}^{2}}{3}\rho-\dot{\frac{H}{2}}.
\end{equation}
Comparing Eq.(17) with the standard Friedmann equation $H^{2}=\frac{8\pi L_{p}^{2}}{3}\rho$
in the GR theory, we find that Eq.(17) has an extra term $-\dot{\frac{H}{2}}$
in the right side. Combining Eq.(13) with Eq.(17), we obtain the following
equation
\begin{equation}
\dot{\rho}+3H(\rho+p)=\frac{3}{8\pi L_{p}^{2}}\left(\frac{\ddot{H}}{2}-\frac{\dot{H}^{2}}{H}\right).
\end{equation}
The R.H.S. of the above equation is nonzero, which indicates that
the evolution of the universe is a nonadiabatic process from the thermodynamic
point of view. This is understandable because there exist temperature
and entropy on the Hubble horizon of the universe. The general form
of the modified continuity equation $\dot{\rho}+3H(\rho+p)=g(H,\;\dot{H})$
where $g(H,\;\dot{H})$ is the function of $H$ and $\dot{H}$ has
been introduced in following three models. The first model is the
bulk viscous cosmology\cite{key-38,key-39} where a bulk viscosity
of cosmological fluid is assumed. The function $g(H,\;\dot{H})=9H^{2}\zeta$
in which $\zeta$ is the bulk viscosity is employed to investigate
the accelerated expansion of the universe (See, for example, in Refs.\cite{key-40,key-41,key-42}).
The second model is the energy exchange cosmology in which the exchange
of energy between two fluids is assumed.  For example, the continuity
equation for matter ``m'' is given as $\rho_{m}+3H(\rho_{m}+p_{m})=-\dot{\Lambda}(t)$
when the dynamical cosmological term $\Lambda(t)$ decays linearly
with the Hubble rate $H$\cite{key-43,key-44,key-45,key-46}. The
third model is the nonadiabatic entropic cosmology where the Hubble
horizon is assumed to have an entropy and a temperature, and the function
$g(H,\;\dot{H})$ is $-\frac{3\gamma}{4\pi L_{p}^{2}}H\dot{H}$\cite{key-35}.
However, in this study, the function $g(H,\;\dot{H})$ is $\frac{3}{8\pi L_{p}^{2}}\left(\frac{\ddot{H}}{2}-\frac{\dot{H}^{2}}{H}\right)$
and the modified continuity equation (18) is the result of combination
of Eq.(13) and Eq.(17). That is to say, the derivation of Eq.(18)
is based on emergence of space and thermodynamics of the Hubble horizon.

At the end of this section, we would like to stress that Eq.(13) and
Eq.(17) are the laws of evolution of the universe obtained by us.
The derivation of Eq.(13) is based on the principle of asymptotically
holographic equipartition and the unified first law, while the derivation
of Eq.(17) is based on the energy balance relation on the Hubble horizon.
Therefore, we describe the evolution of the universe in the language
of thermodynamics in this section .

\section*{3. Formulations and explanation of accelerated expansion of the universe}

As usual, the form of the equation of state of the matter is
\begin{equation}
p=\omega\rho,
\end{equation}
where $\omega$ is a parameter which may change over time. Combining
Eq.(13), Eq.(17) with Eq.(19), we obtain the following equation
\begin{equation}
\dot{H}^{2}+\frac{7+3\omega}{2}H^{2}\dot{H}+3(1+\omega)H^{4}=0.
\end{equation}
This equation can be reduced to
\begin{equation}
\left(H^{2}+\frac{1}{2}\dot{H}\right)\left[3(1+\omega)H^{2}+2\dot{H}\right]=0.
\end{equation}
The equation can be divided into two equations $H^{2}+\frac{1}{2}\dot{H}=0$
and $3(1+\omega)H^{2}+2\dot{H}=0$. Now we discuss these equations
and analyze their physical meanings.

First of all, we discuss the solution $H^{2}=-\frac{1}{2}\dot{H}$.
we get the scale factor
\begin{equation}
a(t)\sim t^{1/2}
\end{equation}
under this solution. The deceleration parameter in cosmology is defined
by
\begin{equation}
q\equiv-\left(\frac{\ddot{a}}{aH^{2}}\right)_{t=t_{0}}.
\end{equation}
Substituting Eq.(22) into Eq.(23), we obtain the deceleration parameter
$q=1$ which implies that our universe has a tendency of decelerated
expansion. As we know, there exists attraction between any kinds of
ordinary matter and gravity has a tendency to decelerate the expansion
of the universe decelerate. Therefore, the universe has to have a
tendency to decelerate as long as there exists ordinary matter, and
this solution is a reflection of this property. This solution is consistent
with that of the epoch dominated by the radiation in the GR theory.

Next, we shall analyze the equation $3(1+\omega)H^{2}+2\dot{H}=0$
and find that this equation is the same as the solution of evolution
of the universe in the GR theory. In order for the completeness of
the discussion, we analyze the solution $H^{2}=\begin{cases}
-\frac{2}{3(1+\omega)}\dot{H} & \quad\omega\neq-1\\
constant & \quad\omega=-1
\end{cases}$ and present the nature of evolution of the universe in the following
paragraphs.

The ``Hubble constant'' $H$ is a true constant when $\omega=-1$.
We can get the scale factor $a(t)\sim e^{Ht}$ which is known as the
de Sitter model\cite{key-47,key-48}. In addition, the solution is
also related to inflation at the early stage of the universe. When
$H^{2}=-\alpha\dot{H}$ where $\alpha=\frac{2}{3(1+\omega)}$, we
get the scale factor
\begin{equation}
a(t)\sim t^{\alpha}.
\end{equation}
Substituting Eq.(24) into Eq.(23), we obtain the deceleration parameter
$q=\frac{1}{\alpha}-1$. Our universe is in accelerated expansion
according to the astronomical observation, so the deceleration parameter
$q$ has to be less than 0 which implies $\alpha>1$ and $-1<\omega<-\frac{1}{3}$.
In order to see the nature of evolution of the universe clearly, we
choose several specific values of $\omega$. We choose the following
values:

\begin{center}
\begin{tabular}{|c|c|}
\hline
$\omega$ & $\alpha$\tabularnewline
\hline
\hline
$-0.52$ & $1.4$\tabularnewline
\hline
$-0.67$ & $2.0$\tabularnewline
\hline
$-0.78$ & $3.0$\tabularnewline
\hline
\end{tabular}
\par\end{center}

The simple relation between the age of the universe and the Hubble
constant gives the age of the universe
\begin{equation}
t_{0}=\frac{\alpha}{H_{0}},
\end{equation}
where $H_{0}$ is the current Hubble constant. Substituting the current
Hubble constant $H_{0}=70km\cdot s^{-1}\cdot Mpc^{-1}$ into Eq.(25),
we obtain that the age of the universe is $14\alpha$ billion year.

The luminosity distance $d_{L}$ has been widely used to investigate
the accelerated expansion of the universe and it is given as \cite{key-35,key-49}
\begin{equation}
d_{L}=\frac{c(1+z)}{H_{0}}\int_{1}^{1+z}\frac{dy}{H/H_{0}},
\end{equation}
where the variable $y$ is defined as $y=a_{0}/a$ and $z$ is the
redshift defined by $z+1\equiv y=a_{0}/a$. Substituting $a(t)=t^{\alpha}$
into Eq.(26), we have
\begin{equation}
\frac{H_{0}}{c}d_{L}=\frac{1+z}{1-1/\alpha}\left[(1+z)^{1-1/\alpha}-1\right]
\end{equation}
Substituting $\alpha=3.0$, $\alpha=2.0$ and $\alpha=1.4$ into Eq.(27)
and plotting the dependent relation curves between the luminosity
distance and the redshift factor, we can obtain the above three curves
of FIG.1 respectively.

For the $\Lambda CDM$ model, the luminosity distance of the flat
universe is given by Refs.\cite{key-35,key-50}, its form is
\begin{equation}
(\frac{H_{0}}{c})d_{L}=(1+z)\int_{0}^{z}dz^{'}[(1+z^{'})^{2}(1+\Omega_{m}z^{'})-z^{'}(2+z^{'})\Omega_{\Lambda}]^{-1/2},
\end{equation}
where $\Omega_{m}$ and $\Omega_{\Lambda}$ are the ratios of the
non-relativistic matter and the vacuum energy to the critical energy
density respectively. It has been found that the relation curve between
the luminosity distance and the redshift factor for $\Omega_{m}=0.23$
and $\Omega_{\Lambda}=0.73$ is consistent with the fitting curve
of the data obtained by WMAP\cite{key-51,key-52}, and the curve represented
by Eq.(28) is the curve at the bottom of FIG.1. Analyzing FIG.1, we
find that these curves are well in line with the result of astronomical
observations. In particular, the curve represented by $\alpha=1.4$
agrees very well with the curve fitted by the observed supernova data
(see FIG.2) and the age of the universe will be $19.6$ billion
year if the universe has been evolving in this way.

\begin{figure}
\centering
\includegraphics[width=8cm]{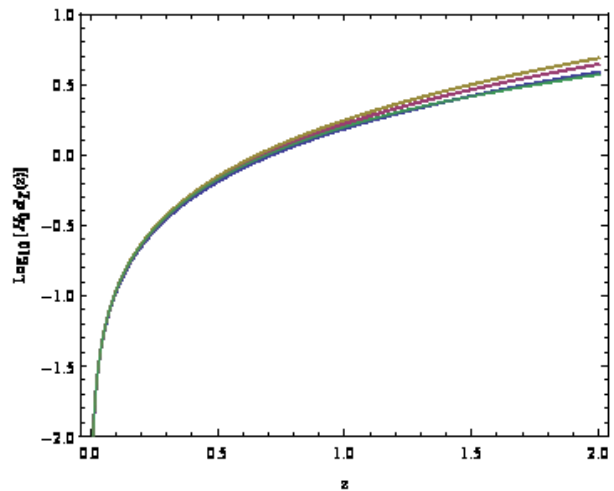}
\caption{(color online). The above three curves correspond to $\alpha=3.0$,
$\alpha=2.0$ and $\alpha=1.4$ from top to bottom respectively, and
the curve at the bottom is a fitting curve of the data obtained by
astronomical observations.}
\end{figure}

\begin{figure}
\centering
\includegraphics[width=8cm]{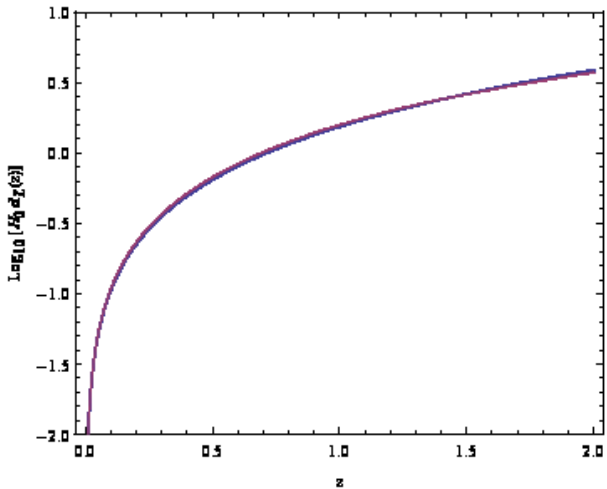}
\caption{(color online). The blue curve (the upper curve) represents
the relation curve between the luminosity distance and the redshift factor
for $\alpha=1.4$, and the red curve (the lower curve) is a fitting
curve of the data obtained by astronomical observations.}
\end{figure}

\section*{4. Comments on the alternative perspective for the evolution of the
universe}

The conventional way of describing the evolution of the universe is
employing the GR theory and cosmological principle. In the GR theory,
the evolution equations of the flat universe are $\frac{\ddot{a}}{a}=H^{2}+\dot{H}=-\frac{4\pi L_{p}^{2}}{3}(\rho+3p)$
and $H^{2}=\frac{8\pi L_{p}^{2}}{3}\rho$. Combining the two equations
with the equation of state of the cosmic matter, we can obtain $3(1+\omega)H^{2}+2\dot{H}=0$
which is only a solution of Eq.(21). Our results are more general
because they can cover the results obtained in the GR theory. There
has always to be the solution $a(t)\sim t^{1/2}$ throughout the evolutionary
history of the universe if the evolution of the universe obeys the
equations Eq.(13) and Eq.(17). We think that this solution is the
evolution law of a possible universe in which the mass of matter equals
exactly the mass of anti-matter. If we only consider the solution
$3(1+\omega)H^{2}+2\dot{H}=0$, then the thermodynamical description
in our model is equivalent to the dynamical description in the GR
theory from the perspective of evolution of the universe. However,
the physical picture is still different, the vacuum energy has a shift
which changes with time in our model while the vacuum energy is constant
in the GR theory. In our models, the entire evolutionary history of
the universe can be described by the radiation dominated period, non-pressure
dominated period and vacuum energy dominated period. The age of the
universe is less than $t_{0}=13.4_{-1.0}^{+1.3}\times10^{9}yr$ indicated
by the data of the supernova\cite{key-2,key-53} if the universe evolves
according to the laws of  the radiation dominated period or non-pressure
dominated period. The age of the universe is greater than $t_{0}=13.4_{-1.0}^{+1.3}\times10^{9}yr$
when the universe evolves according to the laws of  the vacuum energy
dominated period. So the law of evolution of the universe has to change
from a decelerated expansion to an accelerated expansion at a certain
point in time. The result is compatible with astronomical observations\cite{key-5}.

Now we discuss the physical meaning of the temperature of the Hubble
horizon. Jacobson pointed out that the equilibrium thermodynamic
relation $\delta Q=TdS$ is equivalent to the Einstein equation. Moreover,
the first law of thermodynamics $TdS=dE+PdV$ is also equivalent to
the field equations in a very wide class of gravitational theories
as we have pointed out in the introduction. All these show that our
universe can be viewed as an equilibrium system (that is to say, the
evolution of the universe can be viewed as a series of quasistatic
processes) when the temperature is taken to be the temperature of
the horizon from the thermodynamic point of view. Hence, it is reasonable
to employ the temperature of the Hubble horizon as the temperature
of the universe in our model.

\section*{5. Conclusions}

In this paper, we study the accelerated expansion of the universe
based on emergence of space and thermodynamics of the Hubble Horizon.
For this purpose, we derive two evolution equations of the universe from
a thermodynamical point of view. One is the Friedmann acceleration
equation which is derived from emergence of space and thermodynamics
of the Hubble horizon whose temperature is obtained from the unified
first law of thermodynamics. The other is obtained from the energy
balance relation $\rho V_{H}=TS$. In the process of derivation, we
employ the temperature of the Hubble horizon as the temperature of
the universe. This is reasonable because numerous researches have
shown that our universe can be viewed as an equilibrium system from
the thermodynamic point of view.

Combining the two evolution equations and the equation of state of
the cosmic matter, we determine the evolution equation of the FRW
universe. Then we analyze the solutions and discuss their physical meaning.
For the solution $H^{2}=-\frac{1}{2}\dot{H}$, we think that this
is the evolution law of a possible universe in which the mass of matter
equals exactly the mass of anti-matter. For the solution $3(1+\omega)H^{2}+2\dot{H}=0$,
we find that this solution is the same as the solution of the evolution
equation of the universe in the GR theory. In order for the completeness
of the discussion, we plot some relation curves with the parameters
$\alpha=1.4,\;2.0,\;3.0$ of the luminosity distance and the redshift
factor, and find these curves are well in line with the fitting curve
of data of astronomical observations. In particular, the curve represented
by $\alpha=1.4$ agrees very well with the curve fitted by the observed
supernova data. The solutions of the evolution equation of the universe
in our model include the solutions obtained in the GR theory, so it
is more general to describe the evolution of the universe in the thermodynamic
way.

\section*{6. Acknowledgments}
This work is supported by the NNSF of China(Grant No.11375150) and Doctoral Foundation of Zunyi Normal University(Grant No.BS[2016]03). B.S. acknowledges the Science and Technology Foundation of Guizhou Province (Grant No.J[2015]2149). Y.-C.Y. acknowledges the NNSF of China (Grant No.11265017),  the China Postdoctoral Science Foundation (Grant No.2015M571727), and the Guizhou province outstanding youth science and technology talent cultivation object special funds (Grant No.QKHRZ(2013)28).

\bibliographystyle{apsrev4-1}

\end{document}